\documentclass[12pt]{amsart}

\usepackage{amsmath}
 \usepackage{epsfig}
\usepackage{amssymb}
\usepackage{amsfonts}
 \usepackage{latexsym}
 
\newtheorem{Proposition}{Proposition}

  \newtheorem{Remark}[Proposition]{Remark}
  
  \newtheorem{Lemma}[Proposition]{Lemma}
    \newtheorem{Theorem}[Proposition]{Theorem}
  \newtheorem{Note}[Proposition]{Note}

  \makeatletter
\@addtoreset{equation}{section}
\makeatother 
       \def\z{\noindent}

    \def\z{\noindent}  
 
    \def\sqr#1#2{{\vcenter{\vbox{\hrule height .#2pt
                             \hbox{\vrule width .#2pt height#1pt \kern#1pt
                                   \vrule width .#2pt}
                             \hrule height .#2pt}}}}

    \def\drm{\mathrm{d}}

    \def\NN{\mathbb{N}}
    
    \def\RR{\mathbb{R}}
    \def\ZZ{\mathbb{Z}}

\begin{document}

\begin{abstract}
We analyze the time evolution of a one-dimensional quantum system with
zero range potential under time periodic parametric perturbation of
arbitrary strength and frequency. We show that the projection of the
wave function on the bound state vanishes, i.e. the system gets fully
ionized, as time grows indefinitely. 

\vskip -6.5cm

\centerline{{\large \it Dedicated to the memory of J\"urgen Moser}}

\vskip 6.5cm
\smallskip

\end{abstract}

\author{O. Costin, J. L. Lebowitz$^1$ and A. Rokhlenko \\$\mbox{ } $
\\ \Tiny Department of Mathematics\\ Rutgers University 
\\ Piscataway, NJ 08854-8019\\$\mbox{ } $ \\$\mbox{ } $  \\$\mbox{ } $   }


\title{On the complete ionization of a periodically perturbed 
quantum system\\$\mbox{ }$}

\gdef\shorttitle{Complete ionization}
\gdef\shortauthors{O. Costin, J. L. Lebowitz and A. Rokhlenko}
\maketitle

\section{Introduction and  results}

$ $

\z The ionization of atoms\addtocounter{footnote}{1}\footnotetext{Also
  Department of Physics.\\$ $\\ To appear in Proceedings of the CRM
  meeting ``Nonlinear Analysis and Renormalization Group''} subjected to
  external time dependent perturbations is an issue of central
  importance in quantum mechanics which has attracted substantial
  theoretical and experimental interest \cite{[1]}, \cite{[3]}. There
  exists by now a variety of theoretical methods, and a vast amount of
  literature, devoted to the subject.  Beyond the celebrated {}Fermi's
  golden rule, approaches include higher order perturbation theory,
  semi-classical phase-space analysis, {}Floquet theory, complex dilation,
  some exact results for small fields and bounds for large fields and
  numerical integration of the time dependent Schr{\" o}dinger equation
  \cite{[3]}-\cite{[12]}.  Nevertheless there is apparently no complete
  analysis of the ionization of any periodically perturbed model with no
  restrictions on the amplitudes and frequencies of the perturbing
  field. This is not so surprising considering the very complex behavior
  we find in even the most elementary of such systems.

In the present paper we show rigorously the full ionization, in all
ranges of amplitudes and frequencies, of one of the simplest models with
spatial structure which, with a different perturbing potential, is
however frequently used as a model system \cite{[7]}, \cite{[8]},
\cite{[13a]}.  The unperturbed Hamiltonian we consider is

\begin{equation}
  \label{eq:(1)}
  \mathcal{H}_0=-\frac{\hbar^2}{2m}{\frac{\mathrm{d}^2}{\mathrm{d}x^2}}-
g\,\delta
  (x),\ \ g>0,\ \ -\infty<x<\infty.
\end{equation}
$\mathcal{H}_0$ has a single bound state $ u_b(x)=\sqrt{p_0}e^{-p_0|x|},\
p_0=\frac{m}{\hbar^2}g$ with energy $-\hbar \omega_0=-\hbar^2p_0^2/2m$ and a
continuous uniform spectrum on the positive real line, with  generalized
eigenfunctions 
$$u(k,x)=\frac{1}{\sqrt{2\pi}}\left
(e^{ikx}-\frac{p_0}{p_0+i|k|}e^{i|kx|} \right ), \ \ -\infty<k<\infty$$

\z and energies $\hbar^2k^2/2m$.  

Beginning at  $t=0$, we apply a parametric perturbing
potential, i.e. for $t\ge 0$ we have

\begin{equation}
  \label{eq:timedep}
  \mathcal{H}(t)=\mathcal{H}_0 -g\,\eta(t)\delta(x)
\end{equation}
\z and  solve the time  dependent Schr{\"o}dinger
equation for $\psi(x,t)$,
\begin{eqnarray}
  \label{eq:(2)}
 \psi (x,t)=\theta (t)u_b(x)e^{i \omega_0 t}\hskip 4cm\nonumber\\ \hskip
1cm +\int_{-\infty}^
{\infty}\Theta (k,t)u(k,x)e^{-i\frac{\hbar k^2}{2m}t}dk \ \ (t\geq 0)
\end{eqnarray}
with initial values $\theta (0)=1,\ \Theta (k,0)=0$.  This gives the
survival probability $|\theta(t)|^2$, as well as the fraction of ejected
electrons $|\Theta(k,t)|^2 dk$ with (quasi-) momentum in the interval
$dk$.

This problem can be reduced
to the solution of an integral equation \cite{[15]}.   Setting 

\begin{eqnarray}
  \label{eq:(3)}
  &\theta (t)=1+2i\int_0^t Y(s) ds \\
  &\Theta(k,t)= 2|k|/\big[\sqrt{2\pi} (1-i|k|)\big]\int_0^t Y(s)
e^{i(1+k^2)s} ds 
\end{eqnarray}

\z  $Y(t)$ satisfies the integral equation

\begin{equation}
  \label{eq:(5)}
  Y(t)=\eta(t)\left
\{1+\int_0^t [2i+M(t-t')]Y(t') dt'\right \}=\eta(t)\Big(1+(2i+M)*Y\Big)
\end{equation}

\z where $\hbar,2m$ and $\frac{g}{2}$ have been set equal to $1$
(implying $p_0=1$, $\omega_0=1$),
$$ M(s)=\frac{2i}{\pi}\int_0^\infty \frac{u^2e^{-is(1+u^2)}}{1+u^2}du=
\frac{1+i}{2\sqrt{2}\pi}\int_s^\infty\frac{e^{-iu}}{u^{3/2}} du
$$ 

\z and 

$$f*g=\int_0^tf(s)g(t-s)ds$$

\begin{Theorem}\label{T1}
 When $\eta(t)=r\sin\omega t$ 
 the survival probability $|\theta(t)|^2$ tends to
zero as $t\rightarrow\infty$, for all $\omega>0$ and $r\ne 0$.  
\end{Theorem}

\z {\bf Note:} {}For definiteness we assume in the following that $r>0$.

\z The method of proof relies on the properties of the Laplace transform
of $Y$, $y(p)=\mathcal{L}Y(p)=\int_0^{\infty}e^{-pt'}Y(t')dt'$ (note
that $y(p)=\frac{i}{2}(1-p\mathcal{L}\theta)$). In particular we need to
show that $y(p)$ is bounded in the closed right half of the complex $p-$
plane. Before the proof we describe briefly some additional results on
this model system, cf. \cite{[15a]}.

\subsection{{}Further results not proven in the present paper}

({\bf 1}) Theorem \ref{T1}  generalizes to the case when $\eta(t)$
 is a trigonometric polynomial:
\begin{equation}
  \label{eq:(7)}
 \eta(t)=\sum_{j=1}^{K}[A_j\sin(j\omega
  t)+B_j\cos(j\omega t)],
\end{equation}
where we assume $|A_K|+|B_K|\ne 0$.

({\bf 2}) The detailed behavior of the system as a function of $t$,
$\omega$, and $r$ is obtained from the singularities of $y(p)$ in the
complex $p$-plane. We summarize them {\em for small $r$}; below
$\frac{1}{2}<\delta<1$.

At $p=\{i n\omega-i: n\in\ZZ\}$, $y$ has square root branch points and
$y$ is analytic in the right half plane and also in an open neighborhood
${\mathcal{N}}$ of the imaginary axis with cuts through the branch
points. As $|\Im(p)|\rightarrow\infty$ in ${\mathcal{N}}$ we have
$|y(p)|=O(r \omega |p|^{-2})$.  If $|\omega-\frac{1}{n}|> {\rm
const}_nO(r^{2-\delta}) ,\,n\in\ZZ^+$, then for small $ r $ the function
$y$ has a unique pole $p_m$ in each of the strips $ -m\omega>\Im(p)+1\pm
O(r^{2-\delta}) >-m\omega-\omega,\,\ m\in\ZZ$.  $\Re(p_m)$ is strictly
independent of $m$ and gives the exponential decay of $\theta$. After
suitable contour deformation of the inverse Laplace transform, $\theta$
can be (uniquely) written in the form

\begin{eqnarray}
  \label{eq:intform}
 \theta(t)=
e^{-\gamma(r;\omega) t}F_\omega(t)+\sum_{m=-\infty}^\infty
e^{(mi\omega-i)t}h_m(t)&
\end{eqnarray}

\vskip -0.2cm 

\z where $F_\omega$ is periodic of period $2\pi\omega^{-1}$
and 

$$h_m(t)\sim \sum_{j=0}^{\infty}c_{m,j}t^{-3/2-j}\ \
\mbox{as }t\rightarrow\infty, \ \arg(t)\in
\Big(-\frac{\pi}{2}-\epsilon,\frac{\pi}{2}+\epsilon\Big)$$

\z  {\em Not too close
to resonances}, i.e.\ when $|\omega-n^{-1}|>O(r^{2-\delta})$, for all
integer $n$, $|F_{\omega}(t)|=1\pm O(r^2)$ and its Fourier
coefficients decay faster than $r^{|2m|} |m|^{-|m|/2}$. Also, the sum in
(\ref{eq:intform}) does not exceed $O(r^2 t^{-3/2})$ for large $t$, and
the $h_m$ decrease with $m$ faster than $r^{|m|}$.

({\bf 3}) By (\ref{eq:intform}), for times of order $1/\Gamma$ where
$\Gamma=2\Re(\gamma)$, the survival probability for $\omega$ not close
to a resonance decays as $\exp(-\Gamma t)$. This is illustrated in
{}Figure 1 where it is seen that for $r\lesssim 1/2$ the exponential decay
holds up to times at which the survival probability is extremely small,
after which $|\theta(t)|^2=O(t^{-3})$ with many oscillations as
described by (\ref{eq:intform}).  Note the slow decay for $\omega = .8$,
when ionization requires the absorption of two photons.

\begin{figure}
\epsfig{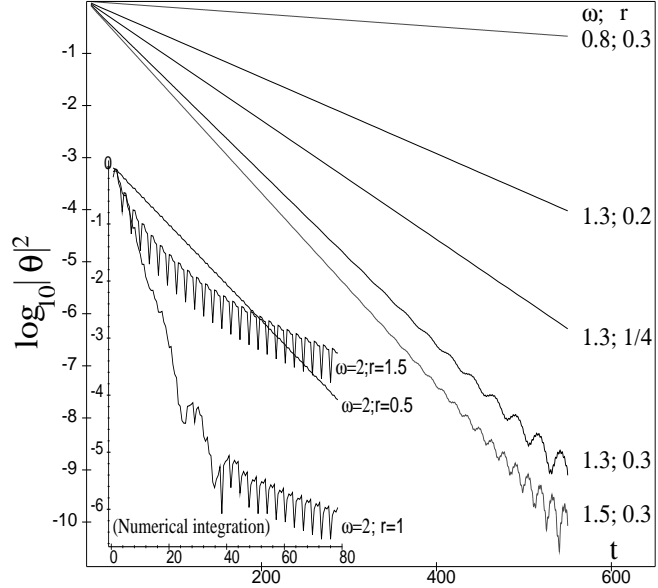}
\caption{Plot of $\log_{10}|\theta(t)|^2$ vs. time in units of
$\omega_0^{-1}$ for several values of $\omega$ and $r$. The main graph
was calculated from (\ref{eq:intform}) and the inset used numerical
integration of (\ref{eq:(5)}).} 
\end{figure}

({\bf 4}) When $r$ is larger the polynomial-oscillatory behavior starts
sooner. Since the amplitude of the late asymptotic terms is $O(r^2)$ for
small $r$, increased $r$ yields higher late time survival
probability. This phenomenon, sometimes referred to as atomic
stabilization  \cite{[12a]}, \cite{[13a]}, can be associated with the
perturbation-induced probability of back-transitions to the well.

({\bf 5}) Using the continued fraction representation (\ref{eq:contfr})
$\Gamma$ can be calculated convergently for any $\omega$ and $r$. 

The limiting behavior for small $r$ of the exponent $\Gamma$ is
described as follows.  Let $n$ be the integer part of $\omega^{-1}+1$ and
assume $\omega^{-1}\notin\NN$. Then we have, for $T>0$ ($t=r^{-2n}T$),

\begin{equation}\label{limit}\hat{\Gamma}=-T^{-1}\lim_{r\rightarrow 0}
\ln\left|\theta (r^{-2n}T)\right|^2
=\frac{2^{-2n+2}\sqrt{n\omega-1}}{\displaystyle{n\omega}\prod_{m<
n}(1-\sqrt{1-m\omega})^2}
\end{equation}

({\bf 6}) The behavior of $\Gamma$ is different at the resonances
$\omega^{-1}\in\NN$. {}For instance, whereas if $\omega$ is not close to
$1$, the scaling of $\Gamma$ implied by (\ref{limit}) is $r^2$ when
$\omega>1$ and $r^4$ when $\frac{1}{2}<\omega<1$, by taking
$\omega-1=r^2/\sqrt{2}$ we find 

$$-T^{-1}\lim_{\begin{subarray}{c} r\rightarrow 0 \\ \omega=1+r^2/\sqrt{2}
\end{subarray}}
\ln\left|\theta (r^{-3}T)\right|^2
=\frac{2^{1/4}}{8}-\frac{2^{3/4}}{16}$$

\section{Proofs of Theorem~\ref{T1}}

\begin{Lemma}\label{L1}
(i) $\mathcal{L}Y$ exists and is analytic in the right half plane
$\mathbb{H}=\{p:\Re(p)>0\}$. {}Furthermore, $y(p)\rightarrow 0$ as
$\Im(p)\rightarrow\pm \infty$ in $\mathbb{H}$.

\z (ii) The function $y(p)$ satisfies (and is determined by) the
functional equation

\begin{align}\label{fnceq}
  y= r \Big(T^{-}-T^{+}\Big)\Big(h_1+h_2 y\Big)
\end{align}

\z with

$$\Big(T^{\pm}f\Big)(p)=f(p\pm i\omega),\ 
  h_1(p)=-\frac{i}{2p}\ \mbox{ and }\ \
  h_2(p)=\frac{1}{2p}\Big(1+\sqrt{1-ip}\Big)$$

\z and by the boundary condition $y(p)\rightarrow 0$ as
$\Im(p)\rightarrow \pm \infty$ in $\mathbb{H}$.

The branch of the square root is such that for $p\in\mathbb{H}$, the
real part of $\sqrt{1-ip}$ is nonnegative and the imaginary part
nonpositive.

\end{Lemma}
\begin{proof} (i) The time evolution of $\psi$ is unitary 
and thus $|\langle \psi\,|\,u_b\rangle|=|\theta(t)|\le 1$. The stated
analyticity is an immediate consequence of the elementary properties of
the Laplace transform \footnote{See also the appendix for a proof of
analyticity for $\Re(p)>p_0$ (all that is required in the subsequent
analysis), relying only on the properties of the convolution
equation.}. The asymptotic behavior follows then from the
Riemann-Lebesgue lemma.

\z (ii) We have in $\mathbb{H}$,

\begin{align}\label{intres}
  \mathcal{L}M=\lim_{a \downarrow 0}\frac{2i}{\pi} \int_0^{\infty}\drm x
  e^{-px}\int_0^{\infty}\frac{u^2e^{-i(x-ia)(1+u^2)}}{1+u^2}\drm
  u\\=\frac{i}{\pi}\int_{-\infty}^{\infty}\frac{u^2}{(1+u^2)(p+i(1+u^2))}\drm u
\end{align}

{}For $\Re(p)>0$ we push the
integration contour through the upper half plane. At the two poles in
the upper half plane $u^2+1$ equals $0$ and $ip$ respectively, so that

\begin{multline}
 \frac{i}{\pi} \int_{-\infty}^{\infty}\frac{u^2}{(1+u^2)(p+i(1+u^2))}\drm
  u\\=\frac{i}{\pi}
\left(
\frac{(-1)}{(2i)(p)}\oint\frac{ds}{s}+\frac{u_0^2}{(ip) (2
  iu_0)}\oint\frac{ds}{s}\right)
=-\frac{i}{p}+\frac{u_0}{p}
\end{multline}

\z where $u_0$ is the root of $p+i(1+u^2)=0$
in the {\em upper} half plane. Thus

\begin{equation}
  \label{eq:e23}
 \mathcal{L}M =
-\frac{i}{p}+\frac{i\sqrt{1-ip}}{p}
\end{equation}

\z with the branch satisfying $\sqrt{1-ip}\rightarrow 1$ as $p\rightarrow 0$ in
$\mathbb{H}$. As $p$ varies in $\mathbb{H}$, $1-ip$ belongs to the lower
half plane $-i\mathbb{H}$ and then $\sqrt{1-ip}$ varies in the fourth
quadrant.

{}For $\Re(p)>0, \omega>0$ we have

\begin{align}
  \mathcal{L}\Big(e^{\pm i\omega}M\Big)=-\frac{i}{p\mp
  i\omega}+\frac{i\sqrt{1-ip\mp\omega}}{p\mp i\omega}\\
(\mbox{with }\ \ \ \sqrt{1-ip-\omega}=-i\sqrt{\omega-1+ip}\ \mbox{ if
}\ \omega>1)\nonumber 
\end{align}

\z and relation (\ref{fnceq}) follows. \end{proof}

\z After the substitution $y(p)=2(\sqrt{1-ip}-1)e^{-\frac{\pi
    p}{2\omega}}v(p)$ we get

\begin{equation}
  \label{eq:rec2}
v(p-i\omega)+v(p+i\omega)=\frac{2}{ r
}(\sqrt{1-ip}-1)v(p)+\frac{i\omega}{\omega^2+p^2}   
\end{equation}

\begin{Remark}
  It is clear that  the functional equation (\ref{eq:rec2}) only
  links the points on one dimensional lattice $\{p+i\ZZ\omega\}$. It
  is convenient to take $p_0$ such that $p=p_0+i n\omega$ with
  $\Re(p_0)=\Re(p)$ and 
\begin{equation}
  \label{eq:norm}
  \Im(p_0)\in [0,\omega)
\end{equation}

\z and write $v(p)=v(p_0+i n\omega)=v_n$ which transforms
(\ref{eq:rec2}) to a recurrence relation:

\begin{equation}
  \label{eq:rec3}
v_{n+1}+v_{n-1}=\frac{2}{ r
  }(\sqrt{1-ip_0+n\omega}-1)v_n+\frac{i\omega}{\omega^2+(p_0+i n\omega)^2}  
\end{equation}

\z where $v_n$ depends parametrically on $p_0$.  It will be seen that the
asymptotic conditions as well as analyticity in $p_0$ determine the
solution of (\ref{eq:rec3}) uniquely.

\end{Remark}

\begin{Remark}
The approach is based on a discrete analog of the Wronskian technique.
The regularity of the bounded solution of (\ref{eq:rec3}) will
be a consequence of the absence of a bounded solution of the homogeneous
equation

\begin{equation}
  \label{eq:homg}
v_{n+1}+v_{n-1}=\frac{2}{ r
  }(\sqrt{1-ip_0+n\omega}-1)v_n=D_n v_n
\end{equation}

\z a
problem which we analyze first.  
\end{Remark}

\begin{Proposition}\label{P1}
{}For $p_0$ satisfying (\ref{eq:norm}) and $\Re(p_0)\ge 0$ (actually for
any $p_0\in \overline{\mathbb{H}}=\mathbb{H}\cup i\RR$) there is no
nonzero solution of (\ref{eq:homg}) such that $v\in l_2(\ZZ)$.

\end{Proposition}

\begin{proof}
  
  To get a contradiction, assume $v\not\equiv 0$ is an $l_2(\ZZ)$ solution of
  (\ref{eq:homg}). Multiplying (\ref{eq:homg}) by
  $\overline{v_n}$, and summing with respect to $n$ from
  $-\infty$ to $+\infty$ we get

\begin{multline}
  \label{eq:sum}
\sum_{n=-\infty}^{\infty}v_{n+1}\overline{v}_n+\sum_{n=-\infty}^{\infty}v_{n-1
}\overline{v}_n
\\=2\sum_{n=-\infty}^{\infty}\Re(v_n\overline{v}_{n+1})
=\sum_{n=-\infty}^{\infty}  \frac{2}{r}\Big(\sqrt{1-ip_0+n\omega}-1\Big)|v|^2_n
\end{multline}

\z {}For $p_0\in \overline{\mathbb{H}}$ the imaginary part of
$\sqrt{1-ip_0+n\omega}$ is nonpositive, by Lemma~\ref{L1}, and is
strictly negative for $n<0$ large enough. Thus if for some such $n$,
$v_n$ is nonzero then the last sum in (\ref{eq:sum}) has a strictly
negative imaginary part, which is impossible since the left side is
real.  If on the other hand $v_n$ is zero when $n$ is large negative,
then solving (\ref{eq:homg}) for $v_{n+1}$ in terms of the $v_{n},
v_{n-1}$ it would follow inductively that $v\equiv 0$, contradicting the
assumption.

\end{proof}

\begin{Lemma}\label{rem1} 
(i) There is, up to multiplicative constants, a unique pair of solutions
$v^+$ and $v^-$ of
(\ref{eq:homg}) such that $v^\pm_n\rightarrow 0$ as $n\rightarrow\pm\infty$ (
respectively). These solutions are related to
convergent continued fractions representations:

\begin{equation}
  \label{eq:contfr} v^{\pm}_{n\mp1}/v^{\pm}_n=: \frac{1}{\rho^{\pm}_n}
=D_n-\frac{1}{D_{n\pm 1}-{\displaystyle \frac{1}{D_{n\pm 2}}}\cdots}
\end{equation}

\z (ii) We have the following estimates

\begin{equation}
  \label{eq:asymptRatio}
\frac{1}{\rho_n^{\pm}}=\frac{1}{\tilde{\rho}_n^{\pm}}+O(n^{-3/2})\ \
(n\rightarrow \pm\infty) 
\end{equation}

\z 
where

\begin{align}
  \label{eq:asymptrho}
&\frac{1}{\tilde{\rho}_n^{+}}=\frac{2}{ r }\sqrt{n\omega}-\frac{2}{ r
}-\frac{ r ^2-2+2ip_0}{2 r 
\sqrt{n\omega}}-\frac{r}{2\omega n}  \ \ \ (n>0)\\
&\frac{1}{\tilde{\rho}_n^-}=
-\frac{2i}{ r }\sqrt{|n|\omega}-\frac{2}{ r }+\frac{(2- r ^2)i+2p_0}{2 r
  \sqrt{|n|\omega}}+\frac{r}{2\omega|n|}\ \ \ (n<0)\nonumber
\end{align}

\z Let $ \tilde{v}^{\pm}_n$ be solutions of the one step recurrences
$\tilde{v}^{\pm}_{n}=\tilde{v}^{\pm}_{n\mp 1}\tilde{\rho}^{\pm}_{n}$. Then

\begin{multline}
  \label{eq:aa}
  \ln \tilde{v}^+_n=-\frac{1}{2}{n\ln
  n}+n\ln\left(\frac{r}{2}\sqrt{\frac{e}{\omega}}\right)
\\+2\sqrt{\frac{n}{\omega}}+\left(\frac{2ip_0+r^2+\omega}{4\omega}\right)\ln
n+o(1) 
\ \ (n\rightarrow\infty)
 \end{multline}
 
\z and

\begin{multline}
  \label{eq:asV2}
\ln( \tilde{v}^-_n)=-\frac{1}{2}{|n|\ln
  |n|}+|n|\ln\left(\frac{r}{2}\sqrt{\frac{e}{\omega}}\right)+i\pi |n|
-2i\sqrt{|n|/\omega}\\+\left(\frac{2ip_0+r^2+\omega}{4\omega}\right)\ln
|n|+o(1) 
\ \ (n\rightarrow -\infty)
\end{multline}

 \z and, for some constants $K^{\pm}$,

\begin{equation}
  \label{eq:asV}
\ln( v^\pm_n)=\ln( \tilde{v}^\pm_n)+K^{\pm}+o(1)
\end{equation}
\z ($v_n^{\pm}$ decay roughly as $1/\sqrt{|n|!}$ for
$n\rightarrow\pm\infty$, respectively).

\smallskip

\z (iii) Two special solutions of (\ref{eq:homg}), $v^+$ and $v^-$, are
well defined by:

\begin{equation}
  \label{eq:defV}
  v^+_n=\tilde{v}^+_n\prod_{j\ge
    n+1}\frac{\tilde{\rho}_j^+}{\rho_j^+}\mbox{ for $n>N$}, \mbox{ and
      }v^-_n=\tilde{v}^-_n\prod_{j\le n-1}
\frac{\tilde{\rho}_j^-}{\rho_j^-}\mbox{ for $n<-N$} 
\end{equation}
\smallskip

\z if $N$ is sufficiently large (this amounts to making a convenient
choice of the free multiplicative constant in (i)). These functions do
not depend on $N$.  $v^+$ and $v^-$ are linearly independent for $p_0\in
\overline{\mathbb{H}}$: their discrete Wronskian, defined by
$W(v^+,v^-)_n=  v^{+}_{n}v^{-}_{n+1}-v^{-}_{n}v^{+}_{n+1}$, satisfies

\begin{equation}
  \label{eq:Wron}
 W(v^+,v^-)=const\ne 0
\end{equation}

\z As functions of parameters, $v^{\pm}$ and $W(v^+,v^-)$ are
 analytic in $p_0\in \mathbb{H}$. If
$\omega\notin\{0,n^{-1}:n\in\NN\}$ then $v^{\pm}$ and $ W(v^+,v^-) $ are
analytic in some neighborhood of $p_0=0$ as well. {}For any $\omega>0$,
$v^{\pm}_n$ are Lipschitz continuous of exponent at least $1/2$ in
$p_0$, for $p_0\in\RR$.
\end{Lemma}

\begin{proof}
  (i) We look at $v^+$, the case of $v^-$ being similar. Dropping the
$^+$ superscript we have from (\ref{eq:homg})

\begin{equation}
  \label{eq:rec2oR}
\rho_{n}=\frac{1}{D_n-\rho_{n+1}}
\end{equation}

\z To find the analytic properties of the solution $\rho_n$ it is
convenient to regard (\ref{eq:rec2oR}) as a contractive equation in the
space $\ell^{\infty}(S_N)$ of sequences $\{\rho_j\}_{j>N}$ in the norm
$\|\rho\|_\infty=\sup_{j>N}|\rho_j|$. Let $N$ be large.  The map
$J:S_N\mapsto S_N$ defined by

\begin{equation}
  \label{eq:defJ}
J(\rho)_n= \frac{1}{D_n-\rho_{n+1}}  
\end{equation}

\z depends analytically on $p_0\in\mathbb{H}$ and is Lipschitz
continuous of exponent at least $1/2$ if $\Re(p_0) \ge 0$. In addition, if
$\|\rho_j\|_\infty\le 1$ we have for sufficiently large $N=N(p_0,\omega,r)$

\begin{equation}
  \label{eq:Jinvar}
  \|J(\rho)\|_\infty\le 
\frac{1}{\frac{2}{r}\left(\sqrt{|N\omega|-1-|p_0|}-1\right)-1}<
\frac{|r|}{|\omega|^{1/2}}\frac{1}{\sqrt{N}}
\end{equation}

\z  Similarly,

\begin{equation}
  \label{eq:Jcontrac}
  \|J(\rho)-J(\rho')\|_\infty\le 
\frac{\|\rho-\rho'\|_\infty}{\left[\frac{2}{r}\left(\sqrt{|N\omega|-1-|p_0|}-1
\right)-1\right]^2}<
\frac{|r|^2}{N|\omega|}\|\rho-\rho'\|_\infty
\end{equation}

\z for sufficiently large $N$ which shows that $J$ is contractive in the
unit ball in $\ell^{\infty}(S_N)$. Thus, equation (\ref{eq:defJ}) has a
unique solution in $S_N$, which depends analytically on
$p_0\in\mathbb{H}$ and is Lipschitz continuous of exponent at least
$1/2$ if $\Re(p_0)\ge 0$. This also implies the convergence of
(\ref{eq:contfr}).

Note that given $\mathcal{K}_1\subset \overline{\mathbb{H}}$ and
$\mathcal{K}_2\subset \RR^+$ both compact, $N$ can be chosen the same
for all $p_0\in \mathcal{K}_1$ and $r\in \mathcal{K}_2$.

\z (ii) {}From (\ref{eq:Jinvar}) it is seen that $|\rho_j|=O(j^{-1/2})$ for
large $j$. Thus, we may write, for large $j$,

\begin{equation}
  \label{eq:appr}
  \frac{1}{\rho^+_j}=\displaystyle D_j-\frac{1}{D_{j+1}-{\displaystyle
\frac{1}{D_{j+2}+O(j^{-1/2})}}} 
\end{equation}

\z The estimates (\ref{eq:asymptRatio}) now follow by a straightforward
calculation. Since $\ln v_n^+=\ln v_{N}^++\sum_{j=N+1}^n\ln\rho^+_j$, the
estimates follow from (\ref{eq:appr}) and the Euler-Maclaurin summation
formula.

\z (iii) As before, we only need to look at $v^+$. We take two compact
sets $\mathcal{K}_1$ and $\mathcal{K}_2$, and choose $N$ as in the note
at the end of the proof of (i). Taking the log in the definition
(\ref{eq:defV}), the infinite sums are absolutely convergent.  By
standard measure theory, $v^+_n$ has the same analyticity properties in
the interior of $\mathcal{K}_1\times \mathcal{K}_2$ and Lipschitz
continuity in $\mathcal{K}_1\times \mathcal{K}_2$ as those of
$\rho^+$, when $n>N$.  Now, (\ref{eq:homg}) easily implies that
the same is true for $n\le N$ as well.

 If $f_n$ and $g_n$ are solutions of (\ref{eq:homg}) then $(
g_{n+1}+g_{n-1})f_n-(f_{n+1}+f_{n-1})g_n=0$ and thus $W_n(f,g)=f_n
g_{n+1}-g_n f_{n+1}=const.$ Thus, if $W_n(f,g)=0$ for some $n$ then
$W_n\equiv 0$ and $f\equiv const~~ g$.  The smoothness properties follow
from the proof of (iii).

 \end{proof}

\begin{Proposition}
  \label{analytic}
  There exists a unique solution of (\ref{eq:rec2}) which is bounded as
  $\Im(p)\rightarrow \pm\infty$ in $\mathbb{H}$.  This solution is
  analytic in $p\in \mathbb{H}$, and $v(p)=O(p^{-2})$ as
  $\Im(p)\rightarrow \pm\infty, p\in \mathbb{H}$.
  
\end{Proposition}

\begin{proof}  By analyticity and continuity $W(v^+,v^-)$ does not vanish
  for any $p\in \mathbb{H}$ and $r>0, \omega>0$. By Lemma~\ref{expEstim}
  the function $v$ defined through $v(p_0+in\omega)=f_n$, where

 \begin{equation}
   \label{eq:fn}
   f_n:=W(v^+,v^-)^{-1}\left(v^+_n\sum_{l=-\infty}^{n-1}v^-_l
H_l+v^-_n\sum_{l=n}^{\infty}v^+_l H_l\right) 
 \end{equation}
 \z and

 \begin{equation}
   \label{eq:eqHn}
   H_n=\frac{i\omega\exp\left(\frac{\pi
      p_n}{2\omega}\right)}{p_n^2+\omega^2}
 \end{equation}

 \z has the required properties. Since no
 solution of the homogeneous equation is bounded on $\ZZ$, $v$ is the
 unique solution with the desired properties.
\end{proof}

\begin{Note}
  \label{N2}
The link between $y$ and $f_n$ is

\begin{equation}
  \label{eq:link}
  y(p)=2(\sqrt{1-ip}-1)e^{-\frac{\pi p}{2\omega}}f_n;\ \ \mbox{ \rm
for }p=p_0+i n\omega 
\end{equation}
\end{Note}
 \begin{Proposition}\label{P:an1}
   The function $y(p)$ is analytic in the right half plane, Lipschitz 
continuous of exponent at least $1/2$
   on the imaginary axis and $\lim_{p\rightarrow 0}y(p)=i/2$.
 \end{Proposition}

 \begin{proof} Since $W$ is analytic in $\mathbb{H}$, continuous and
   nonzero in $\overline{\mathbb{H}}$, $W$ is bounded below in compact
   sets in $\overline{\mathbb{H}}$. Then, the smoothness properties of
   $y$ derive easily from those of $q_n:= W(v^+,v^-)f_n$ on which we
   concentrate now.

\z {\bf (a)} {}For $n\ge 2$ we write, using 
(\ref{eq:eqHn}),

\begin{multline}
  \label{eq:fnreg}
  q_n=v^+_n\sum_{\begin{subarray}{c}{l=-\infty} \cr l\ne
     \pm 1\end{subarray}}^{n-1}v^-_l H_l+v^-_n\sum_{l=n}^{\infty}v^+_l
  H_l\\+v^+_n i\omega e^{\frac{\pi p_0}{2\omega}}\left(
\frac{-iv^{-}_{-1}}{p_0(p_0-2i\omega)}+\frac{iv^{-}_{1}}{p_0(p_0+2i\omega)}
\right)
\end{multline}

\z The last term in parenthesis can be rewritten, using also 
(\ref{eq:homg}), as 

\begin{multline}
  \label{eq:prel}
  \frac{i(v_1^--v_{-1}^-)}{p_0^2+4\omega^2}+\frac{2\omega}{p_0^2+4\omega^2}
\left(\frac{v_1^-+v_{-1}^-}{p_0}\right)\\=
 \frac{i(v_1^--v_{-1}^-)}{p_0^2+4\omega^2}+\frac{4\omega}{r(p_0^2+4\omega^2)}
\frac{\sqrt{1-ip_0}-1}{p_0}v^-_0
\end{multline}

\z Thus we see that $q_{n}$ is continuous as $\Re(p_0)\rightarrow 0$ and
$\Im(p_0)\in [0,\omega)$ [cf. (\ref{eq:norm})], if $n\ge 2$. A very
similar calculation shows the continuity of $q_n$ if $n\le -1$.

\z {\bf (b)} By part (a), $y(p)$ is continuous as $\Re(p)\downarrow 0$
with $\Im(p)\ge 2$ or $\Im(p)<0$. Now, (\ref{fnceq}) written in the form

   \begin{multline}
     \label{eq:rely0}
     r p h_2(p)y(p)=rp\Big(h_1(p+2i\omega)-h_1(p)\Big)\\+rp\Big(y(p+i\omega)
+h_2(p+2i\omega)y(p+2i\omega)\Big)
   \end{multline}
   
   \z shows that $y(p)$ is Lipschitz continuous as $\Re(p)\downarrow 0$
   if $\Im(p)>-2$ thus for all $\Im(p)$. The value of $y(0)$ is easily
   calculated using (\ref{eq:rely0}).
 \end{proof}

 \begin{Proposition}
   $1+2i\lim_{x\rightarrow\infty}\int_0^{x}Y(s)ds=0$.
 \end{Proposition}

 \begin{proof}
 Indeed,   
 
 \begin{multline}
   2\pi
   i\int_0^{\infty}Y(s)ds\\=\lim_{x\rightarrow\infty}\lim_{\delta\rightarrow
     0^+}\left(\int_{-i\infty}^{-i\delta}+\int_{i\delta}^{i\infty}\right)
   \frac{e^{xp}}{p}(i/2+(y(p)-i/2))dp=-\pi
 \end{multline}
\end{proof}

\section{Appendix}

\begin{Lemma}
\label{expEstim}

Equation (\ref{eq:(5)}) has a unique solution $Y\in L^1_{loc}(\RR^+)$ and
$|Y(x)|< K e^{C x}$ for some $K\in\RR^+$ and $C \in\RR$.

\end{Lemma}

\begin{proof}
  Consider $L^1_{loc}[0,A]$ endowed with the norm $\|F\|_{\nu}:=\int_0^A
  |F(s)|e^{-\nu s}ds$, where $\nu>0$. If $f$ is continuous and
  $F,G\in L^1_{loc}[0,A]$, a straightforward calculation shows that

  \begin{align}
    \label{eq:properties}
\|fF\|_{\nu}<\|F\|_\nu\sup_{[0,A]}|f|
\\
\|F*G\|_{\nu}<\|F\|_\nu\|G\|_\nu
\\
\|F\|_\nu\rightarrow 0\ \mbox{ as }\ \nu\rightarrow\infty
  \end{align}
  
  \z where the last relation follows from the Riemann-Lebesgue lemma.

\z The integral equation (\ref{eq:(5)}) can be written as 

\begin{equation}
  \label{eq:eqa}
  Y=r\eta+\mathcal{J}Y \mbox{ where } \ \ \ \mathcal{J} F:=r \eta(2i+M)*F
\end{equation}
  Since $M$ is locally in $L^1$ and bounded for large $x$ it is clear
  that for large enough $C_2$, and for any $A$, (\ref{eq:(5)}) is
  contractive if $\nu>C_2$.
\end{proof}

\textbf{Acknowledgments.} The authors would like to thank  
A. Soffer and M. Weinstein for interesting discussions and suggestions. 
 Work of O. C. was supported by NSF Grant 9704968, that of J.
L. L. and A. R. by AFOSR Grant F49620-98-1-0207 and NSF Grant DMR-9813268.

\end{document}